\documentclass[a4paper, headsepline=on, bibtotoc, 12pt]{scrbook}
\usepackage[UKenglish]{babel}
\usepackage[T1]{fontenc}
\usepackage[latin9]{inputenc} 
\usepackage[left=2.5cm, right=2.5cm, top=4.5cm, bottom=4.5cm]{geometry} 
\usepackage{parskip} 
\usepackage{graphicx} 
\usepackage{color} 
\usepackage{booktabs} 
\usepackage{enumitem} 
\usepackage[labelfont=bf]{caption} 
\usepackage[normalem]{ulem} 
\usepackage{units} 
\usepackage{ziffer} 
\usepackage{braket}

\newcommand{\changefont}[3]{\fontfamily{#1} \fontseries{#2} \fontshape{#3} \selectfont} 
\usepackage[sc]{mathpazo}

\usepackage{amsmath}
\numberwithin{equation}{chapter}
\newtheorem{D}{Definition}[chapter]
\newtheorem{lem}{Lemma}[chapter]
\newtheorem{theo}{Theorem}[chapter]
\newcommand{\x}{\times}

\author{Ramona Wolf\\
 \small Matrikelnummer: 2932650\\\small Fach: B.Sc. Physik
}

\usepackage{bm}

\usepackage{textcomp}

\usepackage{float}

\usepackage{rotating}

\usepackage{longtable}

\numberwithin{figure}{chapter}
\numberwithin{table}{chapter}

\usepackage{titlesec}
\renewcommand{\thechapter}{\arabic{chapter}}
\titleformat{\chapter}[display]
{\bfseries\huge}
{\filleft\MakeUppercase{\chaptertitlename} \Huge\thechapter}
{2ex}
{\titlerule
\vspace{0ex}%
\filright}
[\vspace{1ex}%
\titlerule]
\titlespacing*{\chapter} {0pt}{-50pt}{40pt}

\usepackage[automark]{scrpage2}
\setheadsepline{0.4pt}
\setfootsepline{0.4pt}
\usepackage{pxfonts}
\usepackage{hyperref} 

\begin{document}
\changefont{ppl}{m}{n} 

\renewcommand{\mu}{\muup}



\begin{titlepage}

	\begin{figure}[H]
			\centering
			\includegraphics[width=1\textwidth]{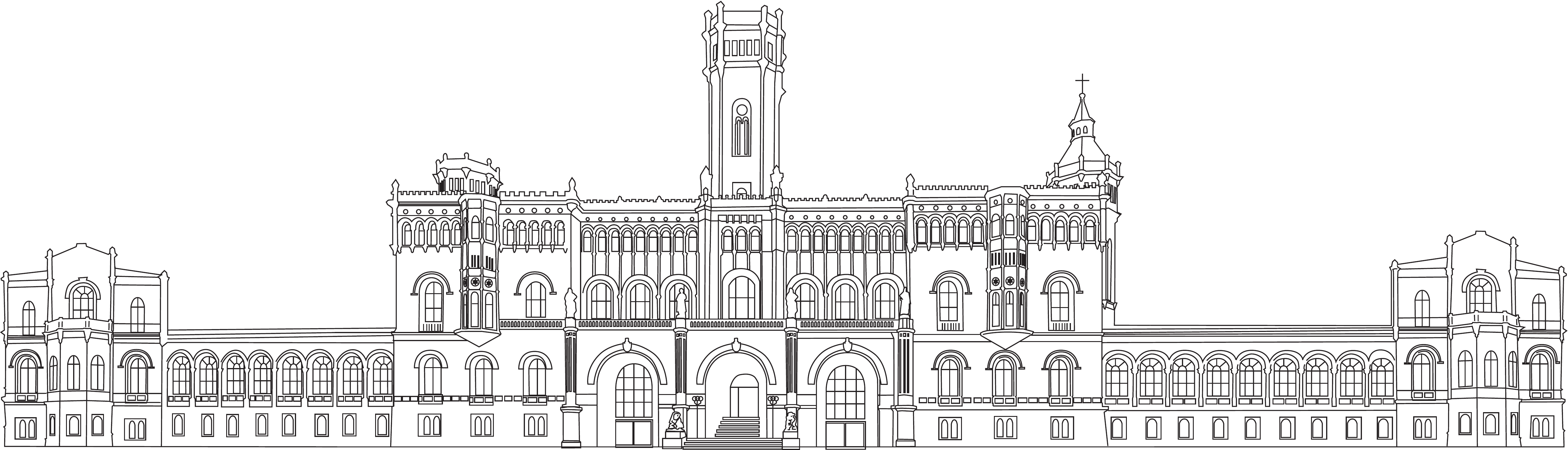}
	\end{figure}
				
\begin{center}

\large
Gottfried Wilhelm Leibniz Universität Hannover \\[0.2cm]
Fakultät für Mathematik und Physik \\[1.5cm]

\noindent\rule{\textwidth}{2pt}
\\
[0.5cm]
\Huge \textbf{Quantum Key Distribution in the Non-Asymptotic Regime} 
\noindent\rule{\textwidth}{2pt}\\
[0.5cm]
\Large
Bachelor Thesis  \\[4.cm]
\large
\textbf{Author:} Ramona Wolf
\\[0.2cm]
\textbf{Supervisor:} Prof. Dr. T. Osborne\\[0.2cm]
\textbf{Co-Supervisor:} Dr. C. Morgan\\[0.2cm]
\text{Hannover, 05.11.2015}
%


\end{center}

\end{titlepage}


\clearpage



\chapter*{Abstract}

How can we generate a key to encrypt messages between two honest parties that an eavesdropper does not gain any information about? The purpose of this thesis is to give an overview over the current state of the prepare and measure quantum key distribution protocol and its security proof with a specific focus on the non-asymptotic case. Therefor, we present the different steps of the protocol and retrace them with the help of the BB84-protocol. Afterwards, the security proof of this protocol is introduced and in the end, we have a closer look at the quantum capacity and calculate it for the amplitude damping channel.
\tableofcontents
\listoffigures

\clearpage





\chapter{Introduction}

The requirement of secure communication is becoming more and more important. Therefore, information theorists work on finding a way to make the exchange of information secure from adversaries. At this point, quantum mechanics has several advantages over classical information theory since it is not a deterministic theory. As Artur Ekert, who is known as one of the inventors of quantum cryptography, said in 1991, ''if nature cannot predict the result, neither can an eavesdropper``.\\
\\
Using the properties of quantum mechanics, information theorists found various applications of quantum information theory that would not be possible with classical information theory. One example is quantum money, a design of bank notes that makes them impossible to forge, proposed by Stephen Wiesner in 1970. Another application is building a quantum computer and finding algorithms where quantum computers outperform classical computers.\\
\\
In this thesis, we are going to investigate another application of quantum information: quantum key distribution. Quantum key distribution is the process of creating a key between two honest parties that an eavesdropper does not gain any knowledge of. This key can later be used to encrypt messages and ensures a secret communication. The goal of this thesis is to get an up-to-date overview of the prepare and measure quantum key distribution protocol (in contrast to the entanglement based protocol) as well as the security analysis with a specific focus on the non-asymptotic case. \\
\\
Therefore, after stating the required mathematical preliminaries in chapter~\ref{Preliminaries}, we introduce the field of entropies in chapter~\ref{Entropies} and discuss the interpretation of the different entropies with regard to quantum information theory. In chapter~\ref{QKD protocol}, we finally present the several steps of the quantum key distribution protocol and retrace them with the help of the well-known BB84 protocol, the first quantum cryptography protocol developed by Charles Bennett and Gilles Brassard in 1984 (see \cite{BB84}). We explain the quantum phase, where the quantum states are prepared and measured as well as the classical post-processing that guarantees that the key is correct and secret. After this, in chapter~\ref{Secrecy analysis}, two different approaches to the secrecy analysis of the protocols are introduced: the way it was done in \cite{RIG} and \cite{VSRR}. Here, the analysis of the non-asymptotic case offers some specifics. In the last chapter, we have a closer look at the quantum capacity, which is the highest rate at which quantum information can be communicated over a noisy quantum channel. We calculate this quantity for a specific channel, the amplitude damping channel that models noise in a quantum channel due to energy loss.

\cfoot{Bachelor thesis - Ramona Wolf}
\newpage
\thispagestyle{myheadings}
\markleft{Introduction}
\thispagestyle{scrheadings}
\setheadsepline{0.4pt}
\setfootsepline{0.4pt}
\cfoot{Bachelor thesis - Ramona Wolf}


\chapter{Preliminaries} \label{Preliminaries}
	
In this chapter, we will introduce some definitions and mathematical terms that will be used throughout the thesis.

\section{Notation}

	A Hilbert space on a system $A$ is denoted $\mathcal{H}_A$ and the product space of two Hilbert spaces is $\mathcal{H}_{AB}=\mathcal{H}_A\otimes \mathcal{H}_B$. The set of operators on a Hilbert-space $\mathcal{H}$ is denoted $\mathcal{B}(\mathcal{H})$.
	\\
	A set of quantum states (i.e. density operators) on the Hilbert-space $\mathcal{H}_A$ is defined as follows:
	\begin{D} [Set of normalised states]
		\begin{equation}
		S_=(\mathcal{H}_A):=\{\rho \in \mathcal{P}(\mathcal{H}_A):Tr(\rho)=1\}
		\end{equation}
	where $\mathcal{P}(\mathcal{H}_A)$ is a set of positive semi-definite operators on $\mathcal{H}_A$, i.e. $\rho\ge 0$.
	\end{D}
	For technical reasons, sub-normalised states are sometimes used instead of normalised states.
	\begin{D} [Set of sub-normalised states]
		\begin{equation}
		S_\le(\mathcal{H}_A):=\{\rho \in \mathcal{P}(\mathcal{H}_A):0 \le Tr(\rho)\le 1\}
		\end{equation}
	\end{D}
	Note that these states have no physical interpretation, but can be thought of as normalised states on a larger Hilbert space $\mathcal{H}'_A$ projected onto $\mathcal{H}_A$.

\newpage
\section{Distance measures}

	Distance measures play a crucial role in quantum information theory. They are typically used to measure the success/failure of a protocol, e.g. by comparing input and output states, or to measure the distinguishability of quantum states. Particularly in the non-asymptotic setting, distance measures are needed to define smooth entropies, because they are evaluated on a quantum state $\rho$ over a set of states $\epsilon$-close to $\rho$ (see definition ~\ref{smoothmin}).\\
	\\
	We are especially interested in a metric on $S_\le (\mathcal{H}_A)$ which will turn out to be the purified distance. To define the purified distance, we need another distance measure, the generalised fidelity:
	\begin{D} [Generalised fidelity]
	For $\rho, \tau \in S_\le(\mathcal{H})$, the generalised fidelity between $\rho$ and $\tau$ is
		\begin{equation}
		F(\rho,\tau):=\sup_{\mathcal{H}'}\sup_{\tilde{\rho},\tilde{\tau}\in S_=(\mathcal{H}')} ||\sqrt{\tilde{\rho}}\sqrt{\tilde{\tau}}||_1
		\end{equation}
	where the supremum is taken over all embeddings $V$ of $\mathcal{H}$ into $\mathcal{H}'$, and all normalised states $\tilde{\rho},\tilde{\tau}\in S_=(\mathcal{H}')$ such that $\rho$ and $\tau$ are images of $\tilde{\rho}$ and $\tilde{\tau}$ under $V^\dagger$, i.e. the states satisfy 
		\begin{align*}
		V^\dagger \tilde{\rho}V=\rho\\
		V^\dagger \tilde{\tau} V=\tau.
		\end{align*}
	\end{D}
	The generalised fidelity can also be written in another way, as the following Lemma shows.
	\begin{lem}
	Let $\rho, \tau \in S_\le(\mathcal{H})$. Then, 
		\begin{equation}
		F(\rho,\tau)=F(\hat{\rho},\hat{\tau})=||\sqrt{\rho}\sqrt{\tau}||_1+\sqrt{(1-\text{Tr}(\rho))(1-\text{Tr}(\tau))}
		\end{equation}
	where $\hat{\rho}:=\rho\oplus(1-Tr\rho),\hat{\tau}:=\tau\oplus(1-Tr\tau)$.
	\end{lem}
	Note that if at least one of the states $\rho$ and $\tau$ are normalised states (i.e. $\text{Tr}(\rho)=1$ or $\text{Tr}(\tau)=1$), the traditional fidelity can be recovered.\\
	Now, we can define the purified distance based on the generalised fidelity:
	\begin{D} [Purified distance]
	For $\rho, \tau \in S_\le(\mathcal{H})$ we define the purified distance between $\rho$ and $\tau$ as
		\begin{equation}
		P(\rho, \tau):=\sqrt{1-F(\rho,\tau)^2}.
		\end{equation}
	\end{D}
	The purified distance is a metric on the set of sub-normalised states $S_\le(\mathcal{H})$ (which is shown in \cite{FNAQIT}, i.e.
		\begin{itemize}
		\item $P(\rho,\tau)=0 \Leftrightarrow \rho=\tau$
		\item $P(\rho,\tau)=P(\tau,\rho)$
		\item The triangle inequality is fulfilled: $P(\rho,\tau)\le P(\rho,\sigma)+P(\sigma,\tau)$.
		\end{itemize}
	Next, we introduce the generalised trace distance, which is a generalisation of the trace distance for sub-normalised states.
	\begin{D}[Generalised trace distance]
	For $\rho, \tau \in S_\le(\mathcal{H})$, the generalised trace distance between $\rho$ and $\tau$ is
		\begin{equation}
		D(\rho,\tau):=\max \{Tr\{\rho-\tau\}_+,Tr\{\tau-\rho\}_+\}
		\end{equation}
	where $\{\hspace{8pt}\}_+$ is the projection onto the positive eigenspace.
	\end{D}
	Note that we introduced two complementary ideas of comparing states:
		\begin{itemize}
		\item Fidelity: measures how \textbf{close} two states are
		\item Purified distance, generalised trace distance: measure how \textbf{distinguishable} two states are
		\end{itemize}
	The main advantage of the purified distance is that we can always find extensions and purifications without increasing the distance (for proofs, see \cite{FNAQIT}).
	\begin{D} [Purification]
	A purification of a state $\rho\in\mathcal{H}_A$ is a pure state $\ket{\psi}^{RA}\in\mathcal{H}\otimes\mathcal{H}'$ with
		\begin{equation}
		\ket{\psi}^{RA}=\sum_{x\in\mathcal{X}} \sqrt{P_X(x)}\ket{x}^R\ket{x}^A
		\end{equation}
	such that $\rho^A=Tr_R(\ket{\psi}\bra{\psi}^{RA})$.
	\end{D}
	\begin{theo} [Uhlmann's theorem for purified distance]
	Let $\rho, \tau \in S_\le(\mathcal{H})$, $\dim \mathcal{H}'=\dim \mathcal{H}$ and $\ket{\psi}\in\mathcal{H}\otimes\mathcal{H}'$ be a purification of $\rho$.
	Then, there exists a purification $\ket{\theta}\in\mathcal{H}\otimes\mathcal{H}'$ of $\tau$ such that 
		\begin{equation}
		P(\rho,\tau)=P(\psi,\theta).
		\end{equation}
	\end{theo}
	\begin{theo}
	Let $\rho, \tau \in S_\le(\mathcal{H})$ and $\tilde{\rho}\in S_{\le}(\mathcal{H}\otimes\mathcal{H}')$ be an extension of $\rho$.
	Then, there exists an extension $\tilde{\tau}\in S_{\le}(\mathcal{H}\otimes\mathcal{H}')$ of $\tau$ such that 
		\begin{equation}
		P(\rho,\tau)=P(\tilde{\rho},\tilde{\tau}).
		\end{equation}
	\end{theo}
	
\section{Completely Positive Maps}
	
	In the quantum key distribution protocol, several steps are modelled with completely positive trace preserving maps, which will be defined in the following. The definitions are taken from \cite{LQI}.
	\begin{D}[Linear, self-adjoint map]
	A linear, self-adjoint map $\epsilon$ is a transformation
		\begin{equation}
		\epsilon:\mathcal{B}(\mathcal{H}_A)\rightarrow\mathcal{B}(\mathcal{H}_B)
		\end{equation}
	which
	\begin{itemize}
	\item is linear, i.e.,
		\begin{equation}
		\epsilon(\alpha\mathcal{O}_1+\beta\mathcal{O}_2)=\alpha\epsilon(\mathcal{O}_1)+\beta\epsilon(\mathcal{O}_2)\hspace{20pt} \forall \mathcal{O}_1,\mathcal{O}_2\in\mathcal{B}(\mathcal{H}_A)
		\end{equation}
	where $\alpha,\beta\in\mathbb{C}$,
	\item and maps Hermitian operators onto Hermitian operators, i.e.,
		\begin{equation}
		\epsilon(\mathcal{O}^\dagger)=(\epsilon(\mathcal{O}))^\dagger \hspace{20pt}\forall\mathcal{O}\in\mathcal{B}(\mathcal{H}_A).
		\end{equation}
	\end{itemize}
	\end{D}
	\begin{D}[Trace preserving map]
	A linear map $\epsilon$ is called trace-preserving if
		\begin{equation}
		\text{tr}(\epsilon(\mathcal{O}))=\text{tr}(\mathcal{O})\hspace{20pt}\forall\mathcal{O}\in\mathcal{B}(\mathcal{H}_A).
		\end{equation}
	\end{D}
	\begin{D}[Positive map]
	A linear, self-adjoint map $\epsilon$ is called positive if
		\begin{equation}
		\forall\rho\in\mathcal{B}(\mathcal{H}_A),\hspace{20pt}\rho\ge 0 \hspace{20pt}\Rightarrow\hspace{20pt}\epsilon(\rho)\ge 0.
		\end{equation}
	\end{D}
	\begin{D}[Completely positive map]
	A positive linear map $\epsilon$ is completely positive if for any tensor extension of the form
		\begin{equation}
		\epsilon'=\mathbb{1}_A\otimes\epsilon
		\end{equation}
	where 
		\begin{equation}
		\epsilon':\mathcal{B}(\mathcal{H}_A\otimes\mathcal{H}_B)\rightarrow\mathcal{B}(\mathcal{H}_A\otimes\mathcal{H}_C),
		\end{equation}
	$\epsilon'$ is positive. Here, $\mathbb{1}_A$ is the identity map on $\mathcal{B}(\mathcal{H}_A)$.
	\end{D}

\chapter{Entropies} \label{Entropies}




	Entropies are an important tool in quantum (and classical) information theory. In general, they measure how much uncertainty there is in a state of a physical system.
	In this chapter, we are going to introduce the most important entropies that are used throughout this thesis.

	First, we have a look at entropies in classical information theory and in the asymptotic i.i.d. setting, that means we just consider independent, identically distributed probability distributions, i.e.
		\begin{equation}
		P_{X^n}(x^n)=\prod \limits_{i=1}^n P_X(x_i)
		\end{equation}
	and an infinite number of trials.\\
	\\
	To define a reasonable entropy, it is crucial that it fulfils certain axioms. First, we consider the entropy of events (see \cite{RR}):
		\begin{itemize}
		\item Independence of representation:\\
			An entropy $H(E)$ of an event $E$ only depends on the probability $P(E)$ of that event.
		\item Continuity:\\
			The function $H$ is continuous in the probability measure $P$.
		\item Additivity:\\
			For two independent events $E$ and $E'$, the function $H$ fulfils	$H(E \cap E')=H(E)+H(E')$.\\
			This is reasonable, because two independent events should not have an influence on each other's uncertainty.
		\item Normalization:\\
			$H(E)=1$ for $E$ with $P(E)=\frac{1}{2}$.\\
			This axiom ensures that the uncertainty of an event is maximal if the probability for an event is $\frac{1}{2}$, i.e. we have no information about it.
		\end{itemize}
	The only function that fulfils all these axioms is
		\begin{equation}
		i(P_X(x)):=-\log P_X(x)
		\end{equation}
	which is shown in \cite{RR} (the logarithm is always taken to the base two).
	With these axioms, it is now possible to define an entropy of a random variable.

\section{Shannon entropy}

	The entropy of a random variable $X$ only depends on the probability mass function $P_X$, analogously to the entropy of an event.\\
	The most standard measure in classical information theory is the Shannon entropy. The Shannon entropy of a random variable $X$ quantifies how much information we gain, on average, if we learn the value of $X$ (which is equivalent to the uncertainty that we have about an event before the measurement).\\
	It is defined as follows:
	\begin{D}[Shannon entropy]
		\begin{equation}
		H(X)=\mathbb{E} [i(P_X(x))]=-\sum \limits_{x\in \mathcal{X}} P_X(x)\cdot \log P_X(x)
		\end{equation}
	where $\mathcal{X}$ is the alphabet, i.e. the values that the realisation $x$ of $X$ can assume.
	\label{shannonent}
	\end{D}

	Apart from the axiomatic justification, it is crucial that an entropy has an operational interpretation.
	The operational interpretation of the Shannon entropy is that $H(X)$ is the optimal compression rate of a source or random variable $X$. For the proof, the weak law of large numbers (W.L.L.N.) and the asymptotic equipartition property (A.E.P.) are used, so at this point it is important that we are in the i.i.d. and asymptotic setting (see \cite{SP},\cite{EIT}).
\newpage
\section{Conditional entropy}

\begin{figure}[H]
			\centering
		  	\includegraphics[width=0.8\textwidth]{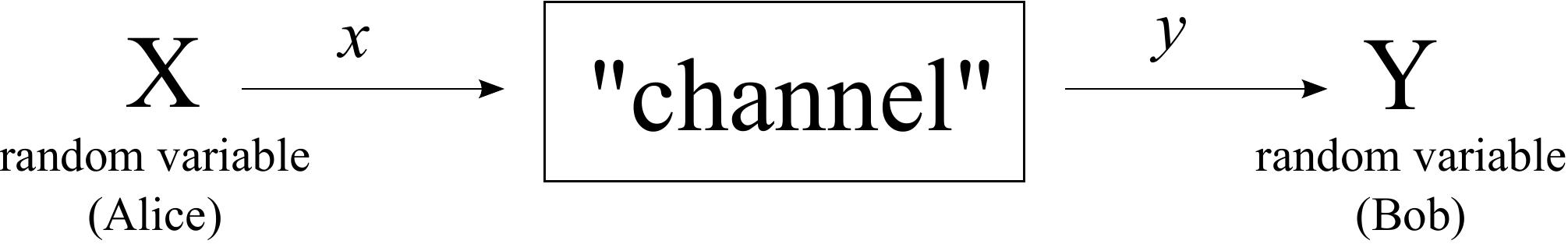}
			\caption{General QIT setting}
			\label{ConEnt}
		\end{figure}

	The general idea in quantum information theory is pictured in figure~\ref{ConEnt}.
	Alice sends a realization $x \in \mathcal{X}$ over a channel to Bob and he receives some information $y \in \mathcal{Y}$.
	Now, Bob would like to learn the value of $X$, knowing $Y$ (i.e. $Y$ is some side information). We would like to quantify Bobs uncertainty about $X$, given $Y$.
	The conditional entropy gives us a value of this uncertainty.\\
	To get an expression for this entropy, we first consider a particular realisation $y \in \mathcal{Y}$. Then, the entropy $H(X|Y=y)$ of a random variable $X$ conditioned on $y$ is:
		\begin{align*}
		H(X|Y=y) &=\mathbb{E}_X[-\log P_{X|Y}(x|y)]\\
		&=-\sum \limits_{x\in \mathcal{X}}P_{X|Y}(x|y) \cdot log(P_{X|Y}(x|y))
		\end{align*}
	With this result, we can find an expression for the conditional entropy:
		\begin{align*}
		H(X|Y) &=\mathbb{E}_{XY}[-\log P_{X|Y}(x|y)]\\
		&=\sum \limits_{y\in \mathcal{Y}}P_Y(y) \cdot H(X|Y=y) \\
		&=-\sum \limits_{y\in \mathcal{Y}}P_Y(y) \sum \limits_{x\in \mathcal{X}}P_{X|Y}(x|y) \cdot \log(P_{X|Y}(x|y))\\
		&=-\sum \limits_{y\in \mathcal{Y}}\sum \limits_{x\in \mathcal{X}}P_{X,Y}(x,y) \cdot \log P_{X|Y}(x|y)
		\end{align*}
	where $P_{X,Y}(x,y)$ is the joint probability mass function (note that if $X$ and $Y$ are independent random variables, $P_{X,Y}(x,y)=P_X(x) \cdot P_Y(y)$).

	Finally, the expression we found for the conditional entropy is
		\begin{equation}
		H(X|Y)=-\sum \limits_{\substack{y\in \mathcal{Y}\\x\in \mathcal{X}}}P_{X,Y}(x,y) \cdot \log P_{X|Y}(x|y)
		\end{equation}

\section{Relative entropy and mutual information}

	Relative entropy is useful to quantify how far one probability distribution $P_{X_1}(x)$ is from another $P_{X_2}(x)$.\\
	Relative entropy is defined as follows:
	\begin{D}[Relative entropy]
		\begin{equation}
		D(P_{X_1}(x)||P_{X_2}(x))=\sum \limits_{x\in \mathcal{X}} P_{X_1}(x)\cdot \log\left(\frac{P_{X_1}(x)}{P_{X_2}(x)}\right)
		\end{equation}
	\end{D}
	This is a useful quantity in its own right, but it can be used to derive other quantities, for example mutual information.\\
	\\
	Mutual information quantifies how much information two random variables X and Y have in common. It is given by the relative entropy between the joint probability $P_{X,Y}(x,y)$ of X and Y and the individual probabilities $P_X(x)$, $P_Y(y)$, i.e.:
	\begin{D}[Mutual information]
		\begin{equation}
		H(X:Y)=\sum \limits_{y\in \mathcal{Y}}\sum \limits_{x\in \mathcal{X}}P_{X,Y}(x,y)\cdot \log\left(\frac{P_{X,Y}(x,y)}{P_X(x)P_Y(y)}\right)
		\end{equation}
	\end{D}
	Figure~\ref{Ent1} shows the relationship between the different entropies (\cite{QCQI}):
		\begin{figure}[H]
			\centering
	  		\includegraphics[width=0.5\textwidth]{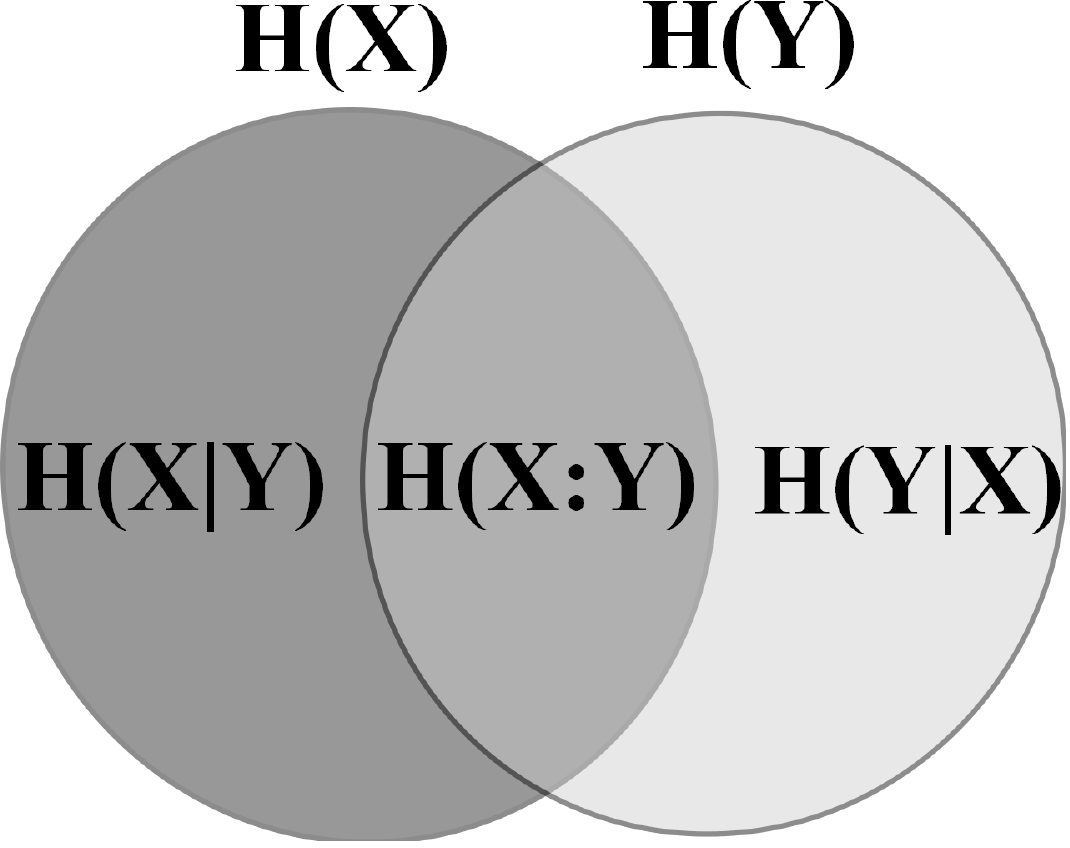}
			\caption{Relationship between entropies}
			\label{Ent1}
		\end{figure}

\section{Classical min- and max-entropy}

	So far, we always considered the asymptotic case. Now, we want to study entropies for the non-asymptotic case and allow a small probability of failure.\\
	\\
	In this case, we cannot use the Shannon entropy anymore, because its operational interpretation depends on the i.i.d./asymptotic setting.
	Reasonable entropies in the non-asymptotic case are the min- and the max-entropy (see \cite{FNAQIT}).
	\begin{D}[Max-entropy]
		\begin{equation}
		H_{\max}(X)_P:=\log\left(\sum \limits_{x\in\mathcal{X}} \sqrt{P_X(x)}\right)^2.
		\end{equation}
	\end{D}
	This entropy can be further refined by smoothing it:\\
	Consider some probability distributions $Q_X$ that are close to $P_X$ and fulfil
		\begin{equation}
		H_{\max}(X)_Q \le H_{\max}(X)_P.
		\end{equation}
	Here, ''close`` means that the $Q_X$ are in a statistical distance $\epsilon$ of $P_X$:
		\begin{equation}
		D(P_X,Q_X):=\frac{1}{2}\sum\limits_{x\in\mathcal{X}}|P_X(x)-Q_X(x)|\le\epsilon.
		\end{equation}
	The smooth max-entropy is now the infimum over all $Q_X$ that are $\epsilon$-close to $P_X$ (in the following denoted by $\approx$):
		\begin{equation}
		H_{\max}^\epsilon(X)_P=\inf_{Q\approx P} H_{\max}(X)_Q.
		\end{equation}
	The smooth max-entropy is used to optimize the Gallagher bound, which is a bound on the minimum code length $m^\epsilon (X)_P$, for which exists an encoder and a decoder that achieve a probability of failure $P\le \epsilon$:
		\begin{equation}
		m^\epsilon (X)_P \le H_{\max}^{\epsilon_1}(X)_P + \log\frac{1}{\epsilon_2}+1
		\end{equation}
	where $\epsilon\le\epsilon_1+\epsilon_2$.
	\begin{D} [Min-entropy]
		\begin{equation}
		H_{\min}(X)_P=\min_{x\in \mathcal{X}}(-\log P_X(x)).
		\end{equation}
	\end{D}
	The classical min-entropy is used for randomness extraction.
	The smoothing part here works the same as the one for the max-entropy.\\
	\\
	For applications, the conditional version of the min-entropy is more interesting. To get an expression for that, we consider the min-entropy with the joint probability distribution:
		\begin{equation}
		H_{\min}(X|Y)_P=-\log \left(\sum\limits_Y P_{XY}(x_*^y,y)\right)
		\end{equation}
	where $x_*^y$ is the $x$ that maximizes $P_{XY}$ for a given $y$.
	The sum in this expression can be written as
		\begin{align*}
		\sum\limits_Y P_{XY}(x_*^y,y) &=\sum\limits_Y P_{Y}(y) \max_{x\in \mathcal{X}} P_X(x)\\
		&=\sum\limits_Y P_{Y}(y)\cdot 2^{\log (\max_{x\in \mathcal{X}}P_X^y(x))}\\
		&=\sum\limits_Y P_{Y}(y)\cdot 2^{-H_{\min}(X)_{P^y}}
		\end{align*}
	With this result, the conditional min-entropy is
		\begin{equation}
		H_{\min}(X|Y)_P=-\log \left(\sum\limits_Y P_{Y}(y)\cdot 2^{-H_{\min}(X)_{P^y}}\right)
		\end{equation}
	where $H_{\min}(X)_{P^y}=log (\max_{x\in \mathcal{X}}P_X^y(x))$ can be considered as the min-entropy of $X$ evaluated for the conditional probability distribution
		\begin{equation}
		P_X^Y(x)=\frac{P_{XY}(x,y)}{P_Y(y)}.
		\end{equation}
	The conditional min-entropy $H_{\min}(X|Y)$ can be interpreted as a guessing probability in tasks such as the following:\\
	Consider an observer with access to Y. We are interested in the probability that this observer guesses X correctly. The optimal strategy is to guess the $x$ with the highest probability conditioned on his observation $y$.	The (average) guessing probability is then given by
		\begin{equation}
		\sum_{Y} P_Y(y)\cdot \max_{x\in\mathcal{X}}P_X^Y(x)=2^{-H_{\min}(X|Y)_P}.
		\end{equation}.

\section{Quantum entropy}

	To find an expression for the entropy in the quantum case, it is crucial to take the two different types of uncertainty into account: The classical uncertainty as well as the quantum uncertainty from the uncertainty principle.
 	Since both types of uncertainty are covered by the density operator, a quantum entropy should be a direct function of the density operator analogously to the classical measurement of uncertainty (see Definition~\ref{shannonent}).
 	\begin{D}[von Neumann entropy]
 	Suppose that Alice prepares some quantum system A in a state $\rho^A$. Then the entropy H(A) of the state is as follows:
 		\begin{equation}
 		H(A):=-Tr\left\{\rho^A\log\rho^A\right\}
 		\end{equation}
 	\end{D}
 	The quantum entropy has several mathematical properties (for proofs, see \cite{QIT}):
 	\begin{itemize}
 	\item Positivity:\\
 		For any density operator $\rho$, the von Neumann entropy is non-negative:
 			\begin{equation}
 			H(\rho)\ge 0.
 			\end{equation}
 	\item Minimum value:
 			\begin{equation}
 			H(\rho)=0\Leftrightarrow\rho \text{ is a pure state.}
 			\end{equation}
 	\item Maximum value:\\
 		When $d$ denotes the dimension of the system, the maximum value is given by $\log d$ and
 			\begin{equation}
 			H(\rho)=\log d\Leftrightarrow\rho \text{ is the maximally mixed state.}
 			\end{equation}
 	\item Concavity:\\
 		The entropy is a concave function in the density operator:
 			\begin{equation}
 			H(\rho)\ge\sum_{x} P_X(x)H(\rho_x)
 			\end{equation}
 		where $\rho=\sum_{x}P_X(x)\rho_x$.
 		This property ensures that entropy never decreases under a mixing operation.
 	\item Unitary Invariance:
 		The von Neumann entropy is invariant under unitary operations:
 			\begin{equation}
 			H(\rho)=H(U\rho U^\dagger)
 			\end{equation}
 	\end{itemize}
 	It is now easy to define different quantum entropies in terms of the von Neumann entropy:
	\begin{D}[Joint quantum entropy]
		\begin{equation}
		H(AB)_\rho=-\text{Tr}\left\{\rho^{AB}\log\rho^{AB}\right\}
		\end{equation}

	\end{D}
	\begin{D}[Conditional von Neumann entropy]
	The conditional quantum entropy $H(A|B)_\rho$ of a bipartite quantum state $\rho^{AB}$ is the difference of the joint quantum entropy $H(AB)_\rho$ and the marginal $H(B)_\rho$:
		\begin{equation}
		H(A|B)_\rho=H(AB)_\rho-H(B)_\rho
		\end{equation}
	\end{D}

\section{Quantum min- and max-entropy}

	Finally, we have a look at the quantum versions of the min- and max-entropy.\\
	What we are actually interested in is the smooth conditional min-entropy. The max-entropy can later be expressed in terms of the min-entropy. Therefore, we first have a look on the definition of the quantum conditional min-entropy:
	\begin{D} [Quantum conditional min-entropy]
	Let $\rho_{AB} \in S_\le (\mathcal{H}_{AB})$ with $\mathcal{H}_{AB}=\mathcal{H}_A \otimes \mathcal{H}_B$. The min-entropy of $A$ conditioned on $B$ of the state $\rho_{AB}$ is
		\begin{equation}
		H_{\min}(A|B)_\rho :=\max_\sigma \sup\left\{\lambda \in \mathbb{R}:\rho_{AB}\le 2^{-\lambda}\mathbb{1}_A\otimes \sigma_B\right\}
		\end{equation}
	where the maximum is taken over all states $\sigma \in S_\le (\mathcal{H}_{B})$.
	\end{D}
	Note that there exists a feasible $\lambda$ only if $\text{supp}\{\rho_B\}\subseteq \text{supp}\{\sigma_B\}$.
	If this assumption is fulfilled, the supremum is achieved by
		\begin{equation}
		\lambda_*=-\log||\sigma_B^{-\frac{1}{2}}\cdot \rho_{AB}\cdot \sigma_B^{-\frac{1}{2}}||_\infty.
		\end{equation}
	The conditional min-entropy can thus be written as
		\begin{equation}
		H_{\min}(A|B)_\rho=\max_\sigma\left\{ -\log||\sigma_B^{-\frac{1}{2}}\cdot \rho_{AB}\cdot \sigma_B^\frac{1}{2}||_\infty\right\}.
		\end{equation}
	The next step is smoothing this entropy. For this step, we will need the definition of an $\epsilon$-ball:
	\begin{D} [$\epsilon$-ball]
	Let $\rho\in S_\le (\mathcal{H})$ and $0\le \epsilon\le \sqrt{Tr(\rho)}$. We define the $\epsilon$-ball of operators on $\mathcal{H}$ around $\rho$ as
		\begin{equation}
		B^\epsilon(\mathcal{H};\rho):=\left\{\tau \in S\le (\mathcal{H}):P(\tau,\rho)\le \epsilon \right\}
		\end{equation}
	where $P(\tau,\rho)$ is the purified distance between $\tau$ and $\rho$.
	\end{D}
	Now, we can use this $\epsilon$-ball for the smoothing parameter of the smooth entropies:
	\begin{D} [Smooth min-entropy]
		\begin{align}
		H_{\min}^\epsilon(A|B)_\rho=\max_{\tilde{\rho}} H_{\min}(A|B)_{\tilde{\rho}}
		\end{align}
	where the optimization is taken over an $\epsilon$-ball of states $\tilde{\rho}$ close to $\rho$.
	\label{smoothmin}
	\end{D}
	Now, we would like to have an expression for $H_{\max}$ as well. Therefore, we use the duality relation between $H_{\min}$ and $H_{\max}$: Let $\rho_{ABC}\in S\le (\mathcal{H}_{ABC})$ be pure. For $\epsilon>0$,
		\begin{equation}
		H_{\min}^\epsilon(A|B)_\rho=-H_{\max}^\epsilon(A|C)_\rho.
		\end{equation}
	The quantum conditional smooth min-entropy is a useful quantity for the secrecy analysis in the non-asymptotic case, see Lemma ~\ref{viereins}.

\section{Entropic uncertainty relation}
\label{eur}
	The uncertainty principle plays a central role in quantum mechanics. The fact that entropic uncertainty relations are indeed desirable was first mentioned by David Deutsch in \cite{EUR}. We present a version of the entropic uncertainty relation that is used in \cite{RIG} to proof the security of the qkd protocol (see theorems ~\ref{theo1} and ~\ref{theo2}).
		\begin{theo}
		Let $\tau_{ABCP}\in\mathcal{S}(ABCP)$ be an arbitrary state with $P$ a classical register. Furthermore, let $\epsilon\in[0,1)$ and let $q$ be a bijective function on $P$ that is a symmetry of $\rho_{ABCP}$ in the sense that $\rho_{ABC,P=p}=\rho_{ABC,P=q(p)}$ for all $p\in P$. Then, we have
			\begin{equation}
			H_{min}^\epsilon(X|CP)_\sigma+H_{max}^\epsilon(X|BP)_\sigma\ge \log \frac{1}{c_q},
			\end{equation}
		where $c_q=\max_{p\in P}\max_{x,z\in X}||F_A^{q(p),x}(F_A^{p,z})^\dagger||_\infty^2$. Here, $\sigma_{XBCP}=\mathcal{M}_{A\rightarrow X|P}(\tau_{ABCP})$ for the map
			\begin{equation}
			\mathcal{M}_{A\rightarrow X|P} [\cdot] = Tr_A \left( \sum_{p \in P} \sum_{x \in X} \ket{x}_X \left( \ket{p}\bra{p}_P \otimes F_A^{p,x} \right) \cdot  \left( \ket{p}\bra{p}_P \otimes F_A^{p,x} \right)^\dagger \bra{x}_X \right)
			\end{equation}
		and any set (indexed by $p\in P$) of generalized measurements $\{F_A ^{p,x}\}_{x\in X}$.
		\end{theo}

\chapter{QKD protocol} \label{QKD protocol}



	In quantum key distribution, two authorized partners (traditionally called Alice and Bob) who want to establish a secret key are connected by two channels: A quantum channel that allows Alice to send quantum signals to Bob and a classical channel, where Alice and Bob can send classical messages back and forth.\\
	An eavesdropper, named Eve, can interact with the signals on the quantum channel, but if she does, the signals are changed due to the no-cloning theorem (see \cite{QIT}). In contrast to the quantum channel, the classical channel is required to be authenticated, that means that Eve can listen to all communication but cannot change the messages.\\
	The general QKD protocol includes several steps. In this part we follow the steps listed in \cite{RIG}. To get an idea of how such a protocol works, we retrace the different steps on the basis of the BB84 protocol (see \cite{QCQI}, \cite{RMP}, and \cite{BB84}). 

\section{Quantum phase}

	The first step is the quantum phase, in which Alice prepares quantum states and sends them to Bob over the quantum channel.
	
	\subsection*{State Preparation}
	For the state preparation, Alice randomly choses a string $r$ of lenght $M$ with $r\in\{0,1\}^M$. The state $\rho_R$ then has the form 
		\begin{equation}
		\rho_R=\frac{1}{2^M}\sum_{r\in\{0,1\}^M}\ket{r}\bra{r}_R
		\end{equation}
	where $\{r\}$ is an orthonormal basis of $R$.\\
	Alice prepares a state by using the map
		\begin{equation}
		\mathcal{P}_{\emptyset\rightarrow A|RS^{\Phi_A}}(\cdot)=\sum_{r,\phi\in\{0,1\}^M}\left(\ket{r}\bra{r}_R\otimes\ket{\phi}\bra{\phi}_{S^{\Phi_A}}\right)\cdot\left(\ket{r}\bra{r}_R\otimes\ket{\phi}\bra{\phi}_{S^{\Phi_A}}\right)\otimes\rho_A^{r,\phi}
		\end{equation}
	where $\rho_A^{r,\phi}=\bigotimes_{i=1}^M\rho_{A_i}^{r_i,\phi_i}$. $A$ is Alice's initial quantum system and $S^{\Phi_A}$ a random seed for choosing the random bit string $\phi$ (for more details about random seeds, see \cite{RIG}).\\
	After applying the map, the resulting state is
		\begin{equation}
		\rho_{RS^{\Phi_A}A}=\frac{1}{4^M}\sum_{r,\phi\in\{0,1\}^M}\ket{r}\bra{r}_R\otimes\ket{\phi}\bra{\phi}_{S^{\Phi_A}}\otimes\rho_A^{r,\phi}
		\end{equation}
	 
	In the BB84 protocol, Alice starts with two strings of random classical bits, $a$ and $b$, each of length $(4+\delta)n$. She encodes these strings as a block of $(4+\delta)n$ qubits:
		\begin{equation} \ket{\psi}=\bigotimes\limits_{i=1}^{(4+\delta)n}\ket{\psi_{a_i b_i}}
		\end{equation}
	where $a_i$ is the $i^{th}$ bit of $a$ and similar for $b$.
	Every qubit is in one of the four states
		\begin{center}
		$\ket{\psi_{00}}=\ket{0}$\\
		$\ket{\psi_{10}}=\ket{1}$\\
		$\ket{\psi_{01}}=\ket{+}=(\ket{0}+\ket{1})/\sqrt{2}$\\
		$\ket{\psi_{11}}=\ket{-}=(\ket{0}-\ket{1})/\sqrt{2}$
		\end{center}
	Note that the basis in which $a_i$ is encoded is determined by the bit $b_i$. If $b_i$ is 0, $a_i$ is encoded in the computational basis, for $b_i=1$ in the Hadamard basis. After the encoding, the qubits are in states which are not all mutually orthogonal, which means it is impossible to distinguish between all of them with certainty without knowing $b$.
	
	\subsection*{State distribution}
	After the state has been prepared, Alice sends it to Bob over the quantum channel $\mathcal{N}:A\rightarrow B$. $B$ is Bob's initial quantum system. Bob receives the state 
		\begin{align}
		\begin{split}
		\rho_{RS^{\Phi_A}B}&=\mathcal{N}_{A\rightarrow B}(\rho_{RS^{\Phi_A}A})\\
		&=\frac{1}{4^M}\sum_{r,\phi\in\{0,1\}^M}\ket{r}\bra{r}_R\otimes\ket{\phi}\bra{\phi}_{S^{\Phi_A}}\otimes\rho_B^{r,\phi}
		\end{split}
		\end{align}
	with $\rho_B^{r,\phi}=\mathcal{N}(\rho_A^{r,\phi})$.
	
	\subsection*{Measurement}
	
	To choose the basis he measures in, Bob uses a random string $\Phi_B$. The outcomes of his measurements are either 0, 1, or $\emptyset$ for inconclusive measurement results, for example due to a photon loss. Bob stores these outcomes in a string $T\in\{0,1,\emptyset\}^M$. The measurement map he uses is
		\begin{equation}
		\mathcal{M}_{B\rightarrow T\Omega|S^{\Phi_B}}(\cdot)=\sum_{\phi\in\{0,1\}^M} \sum_{t\in\{0,1,\emptyset\}^M} \ket{t,\omega}_{TC^\Omega}\left(M_B^{\phi,t}\otimes\ket{\phi}\bra{\phi}_{S^{\Phi_B}}\right)\cdot\left(M_B^{\phi,t}\otimes\ket{\phi}\bra{\phi}_{S^{\Phi_B}}\right)^\dagger\bra{t,\omega}_{TC^\Omega}
		\end{equation}
	where $\omega=\omega(t)$ is the subset of $[M]$ where $t$ takes values in $\{0,1\}$ (i.e. no inconclusive results), namely
		\begin{equation}
		\omega(t)=\{i\in[M]:t_i\ne\emptyset\}.
		\end{equation}
	
	In the BB84 protocol, Bob receives a state $\mathcal{E}(\ket{\psi}\bra{\psi})$, where $\mathcal{E}$ represents the quantum operation on the state due to the noise of the channel and the interaction of an eavesdropper. At this point, Alice, Bob and Eve each have their own states. Note that since Alice is the only one who knows $b$, Eve has no knowledge of what basis she should have measured in to eavesdrop the communication. She can only guess, but if her guess was wrong, she would have disturbed the state received by Bob. \\
	Of course, Bob has no knowledge of $b$ either, but he measures each qubit in one of the bases, determined by a random bit string also of length $(4+\delta)n$, called $b'$. After the measurement, Bob holds a bit string $a'$. 

\section{Sifting phase}

	In the sifting phase (which is optional), Alice and Bob agree to discard some symbols.
	First, Bob publicly announces $S^{\Phi_B}$, i.e. the bases he measured in, and the set $\Omega$ of indices corresponding to conclusive measurement results. After that Alice applies the sifting map 
		\begin{equation}
		\text{sift}:\begin{cases}
		\{0,1\}^M\times\{0,1\}^M\times 2^{[M]} &\rightarrow \Pi_{M,m}\times\{\perp,\not\perp\} \\
		(\Phi_A,\Phi_B,\Omega) &\mapsto (\Sigma,F^{\text{sift}})
		\end{cases}
		\end{equation}
	where $\Sigma$ is the subset of $\Omega$ of cardinality $m$ where $\Phi_A$ and $\Phi_B$ coincide, if such a set exists and the flag $F^{\text{sift}}$ is set to $\not\perp$. Otherwise, it is set to $\perp$ and the protocol aborts.\\
	\\
	In the BB84 protocol, after Alice has announced $b$, she and Bob discard those bits in \{$a'$,$a$\}, that Bob measured in a different basis than Alice, which means that the corresponding bits of $b'$ and $b$ are not equal. After this step, the remaining bits satisfy (in the ideal case) $a'=a$. With high probability, there are 2n bits left ($\delta$ can be chosen sufficiently large).\\
	So far, Alice and Bob have no information if there is an eavesdropper in their communication. To check how much information has been leaked to Eve, Alice selects a random subset of $n$ bits, announces them and compares their values to Bob's values of the checkbits. If more than an acceptable number of them disagree, they abort the protocol.
	
\section{Classical Post-Processing}

	After the sifting phase, the classical post-processing starts. The collected data is estimated and transformed into a secure key (for more details, see \cite{NORM}, \cite{JD}). This step is performed during the BB84 protocol as well, but it depends on the channels that are used and there are often several ways to perform these steps so we will just present how these steps work in general.
	
\subsection*{Parameter estimation}

	Parameter estimation is the first classical post-processing step. 
	In the parameter estimation phase, Alice and Bob want to gain some statistical knowledge about their strings to figure out how many errors have occurred and to get an idea of how much information an eavesdropper may have about their strings.\\
	\\
	For that purpose, Alice sends a small sample of her string to Bob. He compares it to his own string and announces the error rate he observes. From this error rate of the small sample Alice and Bob can get bounds on the number of errors of the whole string (see \cite{NORM}, \cite{SER}).\\
	\\
	If the estimated error rate is too high, Alice and Bob have to abort the protocol (i.e. they set the flag $\perp$), because an eavesdropper may have gained so much information that it is impossible to get a secure key even with a high amount of privacy amplification. Otherwise, they set the flag $\not\perp$ and continue.
	
\subsection*{Information reconciliation}

	After parameter estimation, Alice and Bob know what error rate their strings hold and start to correct these errors. Their aim is to communicate a minimal amount of information to each other to fix the errors.\\
	\\
	Information reconciliation may be a probabilistic procedure, i.e. with high probability it corrects all errors but with a small probability it does not. Hence, Alice and Bob have to check if the procedure was successful and therefore, they need to send a small amount of information over the classical channel. The simplest strategy is to compare hashes of their strings, i.e. they randomly choose a function of a family of two-universal hash functions and apply it to their strings.
		\begin{D}[Two-universal Hashing]
		Let $\mathcal{H}=\{h\}$ be a family of functions from $\mathcal{X}$ to $\mathcal{Z}$. The family $\mathcal{H}$ is said to be two-universal if $Pr[H(x)=H(x')]\le\frac{1}{\mathcal{Z}}$ for any pair of distinct elements $x,x'\in\mathcal{X}$, when $H$ is chosen uniformly random in $\mathcal{H}$.
		\label{two-uni}
		\end{D}
	If their hashes differ, Alice and Bob abort the protocol.
	

\subsection*{Privacy amplification}

	After the information reconciliation step is done, Alice and Bob hold the same strings. Now, they have to remove any information an eavesdropper may have about their shared string. Therefore, they need to send information over the classical channel. The shorter Alice and Bob make their string, the more secure it will be.\\
	\\
	Basically, this step is done by using randomness extractors. A randomness extractor is a function that takes a source of randomness as an input (e.g. a string with a lower bound on its entropy) and a small uniformly random seed. It generates an almost uniformly random output which is longer than the seed. Furthermore, for the application in quantum key distribution, it is crucial that this randomness is extracted with respect to a quantum adversary and additionally the seed and the output string should be independent of each other (so that even if Eve gets access to the seed, she has no information about the output). Altogether, we want a strong randomness extractor against quantum adversaries, that is defined in the following (from \cite{NORM}):
		\begin{D}[Quantum-Proof Strong Randomness Extractor]
		A $(k,\epsilon)$-strong quantum-proof randomness extractor, Ext, is a function from $\{0,1\}^n\times\{0,1\}^d$ to $\{0,1\}^m$ if for all classic-quantum states $\rho_{XE}$ with a classical $X\in\{0,1\}^n$ with min-entropy $H_{min}(X|E)_\rho\ge k$ and a uniform seed $Y\in\{0,1\}^d$ we have
		\begin{equation}
		\frac{1}{2}||\rho_{Ext(X,Y)YE}-\frac{\mathbb{1}}{2^m}\otimes\rho_Y\otimes\rho_E||\le \epsilon
		\end{equation}
		\end{D}
	
	One example of randomness extractors used in QKD are universal hash functions as defined in definition ~\ref{two-uni}. In privacy amplification, Alice and Bob randomly choose a hash function which they both apply on the keys they hold after the information reconciliation step. The fact that universal hash functions make good randomness extractors is ensured by the leftover hashing lemma (\cite{RIG}):
		\begin{lem}[Leftover Hashing Lemma]
		Let $\sigma_{XE'},\tilde{\sigma_{XE'}}\in S_\le (XE')$ be classical-quantum states and let $\mathcal{H}$ be a two-universal family of hash functions from $\{0,1\}^n$ to $\{0,1\}^l$. Moreover, let $\rho_{S^H}=\sum_{h\in\mathcal{H}}\frac{1}{|\mathcal{H}|}\ket{h}\bra{h}_{S^H}$ be fully mixed. Then,
			\begin{equation}
			||\omega_{KS^HE'}-\chi_K\otimes\omega_{S^HE'}||_1\le 2^{-\frac{1}{2}(H_{min}(X|E')_{\tilde{\sigma}}-l)}+2||\sigma_{XE'}-\tilde{\sigma}_{XE'}||_1
			\end{equation}
		where $\chi_K=\frac{1}{2^l}id_K$ is the fully mixed state and $\omega_{KS^HE'}=tr_X(\mathcal{E}_f(\sigma_{XE'}\otimes\rho_{S^H}))$ for the function $f:(x,h)\mapsto h(x)$ that acts on the registers $X$ and $S^H$.
		\end{lem}

\chapter{Secrecy analysis} \label{Secrecy analysis}
		When generating a key in quantum key distribution, not only the protocol itself is important, it is also crucial to analyse the security of the key. Several failure probabilities occur in the process that have to be taken into account (e.g. during information reconciliation). In the non-asymptotic scenario, some special characteristics arise that we are going to investigate in this chapter.\\

\section{Definition and proof of security}

	In general, the security of a protocol (respectively a key) can be expressed in terms of its deviation from a perfect protocol, i.e. one that outputs a uniformly distributed key that is completely independent of the knowledge an eavesdropper has. Therefore, we define a security parameter
		\begin{equation}
		\Delta_{M,k,n,\delta\text{,sift,ec,pa}}:=\sup_{\mathcal{N}_{A\rightarrow BE}}\frac{1}{2}||\text{qkd\_PM}_{M,k,n,\delta,\text{sift,ec,pa}}(\mathcal{N}_{A\rightarrow BE})-\text{qkd\_ideal}_{M,k,n,\delta,\text{ec,pa}}(\mathcal{N}_{A\rightarrow BE})||_1
		\end{equation}
	with the channel as input.\\
	\\
	To be secure, a protocol has to fulfil two aspects: \textbf{correctness} and \textbf{secrecy}. Correctness means, that the probability that the protocol is not aborted (i.e. the flags at parameter estimation and error correction are set to $\not\perp$) though the keys differ is sufficiently low, i.e.
		\begin{equation}
		Pr[K_A\ne K_B\wedge F^{pe}=F^{ec}=\not\perp]\le\epsilon_{ec}.
		\end{equation} 
	Secrecy refers to the fact, that the deviation of the actual key from a perfect key is sufficiently low either:
		\begin{equation}
		Pr[F^{pe}=F^{ec}=\not\perp]\cdot\frac{1}{2}||\omega_{K_A SCFE|F=(\not\perp,\not\perp)}-\chi_{K_A}\otimes\omega_{SCFE|F=(\not\perp,\not\perp)}||_1\le\epsilon_{pa}
		\end{equation}
	Then, the security parameter satisfies $\Delta\le\epsilon_{ec}+\epsilon_{pa}$.\\
	\\
	The following theorems ensure that the protocol fulfils both aspects, correctness and secrecy. The first one refers to correctness:
	\begin{theo}
	For every state $\rho_{ABE}$ and $\omega_{K_A K_B SCFE}=$qkd\_simple$_{k,n,\delta,ec,pa}(\rho_{ABE})$ we have
		\begin{equation}
		Pr[K_A \neq K_B \wedge =(\not\perp,\not\perp)]_\omega\le\epsilon_{ec}:=2^{-t}
		\end{equation}
	\label{theo1}
	\end{theo}
	The second theorem asserts secrecy:
	\begin{theo}
	Define 
		\begin{equation}
		\epsilon_{pa}(\nu):=2^{-\frac{1}{5}(n\log\frac{1}{\bar{c}}-nh(\delta+\nu)-s-t-l)}
		\label{fivefive}
		\end{equation}
	where h is the binary entropy function 
		\begin{equation}
		h:x\mapsto -x\log x-(1-x)\log(1-x).
		\end{equation}
	If $\epsilon(0)<1$, define $\nu_*$ as the unique solution of the equality
		\begin{equation}
		\epsilon_{pa}(\nu)=\exp\left(-\frac{nk^2\nu^2}{2(n+k)(k+1)}\right).
		\end{equation}
	If furthermore, this solution satisfies $\epsilon_{pa}(\nu_*)\le\frac{1}{4}$, then, for every state $\rho_{ABE}$ and $\omega_{K_A K_B SCFE}=$\\
	qkd\_simple$_{k,n,\delta,ec,pa}(\rho_{ABE})$, we have
		\begin{equation}
		Pr[F=(\not\perp,\not\perp)]_\omega\cdot\frac{1}{2}||\omega_{K_A SCFE|F=(\not\perp,\not\perp)}-\chi_{K_A}\otimes\omega_{SCFE|F=(\not\perp,\not\perp)}||_1\le\epsilon_{pa}(\nu_*)
		\end{equation}
	\label{theo2}
	\end{theo}
	In \cite{RIG}, these theorems are proved which leads to the proof of the security of the protocol. Therefore, they first proof the security of the simple and rather unrealistic entanglement based protocol. Then, they make several assumptions to trace the proof of the prepare and measure protocol back to the one of the entanglement based protocol. The heart of their proof is the entropic uncertainty relation that was already mentioned in chapter~\ref{eur}.\\
	\\
	In \cite{VSRR}, they have a slightly different acces to the security of the protocol. They require two properties of a key: First, the characteristic of the non-asymptotic scenario is that this deviation $\epsilon$ from a perfect key is always finite and therefore needs an operational interpretation to make it possible to find reasonable security thresholds. Additionally, another relevant requirement in practical QKD is the composability of the key which guarantees that the generated key can be safely used in applications. The following definition considers both conditions, composability and the operational interpretation of $\epsilon$. 
	\begin{D}[$\epsilon$-secure key]
	For any $\epsilon\ge 0$, a key K is said to be $\epsilon$-secure with respect to an adversary E it the joint state $\rho_{KE}$ satisfies
		\begin{equation}
		\frac{1}{2}||\rho_{KE}-\tau_K\otimes\rho_E||_1\le\epsilon
		\end{equation}
	where $\tau_K$ is a completely mixed state on $K$.
	\label{securekey}
	\end{D}
	The operational interpretation of the parameter $\epsilon$ in this definition is that it can be seen as the maximum probability that $K$ differs from a perfect key, or equivalently as the maximum failure probability of the protocol.\\
	It is also easy to see why this definition is composable: Consider a cryptosystem that uses a perfect key. By replacing the perfect key with an $\epsilon$-secure key, its failure probability only increases by at most $\epsilon$.
	
\section{Non-asymptotic analysis}

	This analysis is based on \cite{VSRR}. Among other aspects, it gives a bound on the final key length $l$, that already appeared in equation ~\ref{fivefive} and an expression for the sifted key rate $r'$ in the non-asymptotic case.\\
	\\
	In the asymptotic case, i.e. the size of the raw key tends to infinity, the quality of a protocol is commonly expressed in terms of the sifted key rate $r'$, defined as
	\begin{D}[sifted key rate]
		\begin{equation}
		r':=\lim\limits_{n\rightarrow\infty}\frac{l(n)}{n}
		\end{equation}
	where $l(n)$ is the number of generated key bits and $n$ is the size of the raw key.
	\end{D}
	Under the assumptions of collective attacks, the sifted key rate can be written in terms of the conditional von Neumann entropy:
		\begin{equation}
		r'=H(X|E)-H(X|Y)
		\end{equation}
	i.e. the sifted key rate is equal to the difference between Eve's uncertainty on the raw key bits $X$ and Bob's uncertainty.
	Multiplying the sifted key rate $r'$ with $\frac{n}{M}$ yields to the key rate per signal $r$. In many schemes in the asymptotic case, $\frac{n}{M}$ can be chosen arbitrary large to 1, so the sifted key rate $r'$ and the key rate per signal $r$ are asymptotically equal.\\
	\\
	However, in the non-asymptotic case, the number of exchanged signals $M$ is always finite. There are several deviations from the asymptotic case, e.g. in parameter estimation or information reconciliation where one needs to find a trade-off between the length of the key and the precision of the post-processing. The aim is to find an expression for the sifted key rate $r'$ analogously to the asymptotic case.\\
	\\
	The first step to derive such an expression is finding a bound on the number of final key bits $l$:
	\begin{lem}
	The protocol described above generates an $\epsilon$-secure key if, for some $\bar{\epsilon}\ge 0$
		\begin{equation}
		l\le H^{\bar{\epsilon}}_{\min}(X^n|E^n)-\text{leak}_{EC}-2\log\frac{1}{2(\epsilon-\bar{\epsilon}-\epsilon_{EC})}
		\end{equation}
	where $leak_{EC}$ is the amount of information leaked during information reconciliation.
	\label{viereins}
	\end{lem}
	By evaluating the smooth min-entropy and taking into account that the states have to be in a set compatible with the statistics from the parameter estimation, we can find an expression for $r'$:
	\begin{lem}
		\begin{equation}
		r'=H_{\xi}(X|E)-(\text{leak}_{EC}+\Delta)/n
		\end{equation}
	with $H_\xi(X|E)=\min_{\sigma_{\bar{X}\bar{E}}\in\Gamma_\xi}H(X|E)$ and $\Delta=2\log\frac{1}{2(\epsilon-\bar{\epsilon}-\epsilon_{EC})}+\frac{7}{\sqrt{n\log(2/(\bar{\epsilon}-\bar{\epsilon}'))}}$ (with the assumption $\bar{\epsilon}>\bar{\epsilon}'$).
	$\Gamma_\xi$ is a set of states compatible with the statistics $\lambda_{(a,b)}$ from parameter estimation.
	\label{lem513}
	\end{lem}
	Several errors from different sources occur in this expression:
	\begin{itemize}
	\item $\epsilon$ is the probability that the key differs from a perfect key
	\item $\epsilon_{EC}$ is the failure probability of the information reconciliation, i.e. that Bob computes a wrong guess for $X^n$
	\item $\bar{\epsilon}$ is the smoothing parameter of the min-entropy (see lemma ~\ref{viereins})
	\item $\bar{\epsilon}'$ is the probability that the set $\Gamma_\xi$ is not compatible with the statistics $\lambda_{(a,b)}$
	\end{itemize}
	Note that $M$, $\epsilon$, leak$_{EC}$ and $\epsilon_{EC}$ are parameters that are implemented in the protocol whereas $n$, $m$, $\bar{\epsilon}$ and $\bar{\epsilon}'$ have to be chosen to maximize the key rate per signal $r=(n/M)r'$ under the constraints $n+m\le M$ and $\epsilon-\epsilon_{EC}>\bar{\epsilon}>\bar{\epsilon}'$.

\chapter{Quantum capacity} \label{Quantum capacity}
		After studying the QKD protocol and its security in detail, we have a look at one important quantity for the communication over quantum channels: the quantum capacity. The quantum capacity is the highest rate at which quantum information can be communicated over a noisy quantum channel. This implies that the requirement for the key rate $r$ for a secure communication is
		\begin{equation}
		r\le\mathcal{Q}(\mathcal{N})
		\end{equation}
	where the error $\epsilon$ tends to $0$ as $n$ tends to infinity. \\
	\\
	In general, the quantum capacity of a quantum channel is the supremum over all achievable rates for quantum communication, i.e. $\sup\{\mathcal{Q}|\mathcal{Q}\text{ is achievable}\}$. It was shown by Lloyd (\cite{Lloyd}) Shor (\cite{Shor}) and Devetak (\cite{Devetak}), that the quantum capacity can be expressed as
		\begin{equation}
		\mathcal{Q}(\mathcal{N})=\lim\limits_{n\rightarrow\infty}\frac{1}{n}\mathcal{Q}^{(1)}(\mathcal{N}^{\otimes n})
		\end{equation}
	with the single-letter expression
		\begin{equation}
		\mathcal{Q}^{(1)}=\max_{\phi^{AA'}}\{I(A\rangle B)_\rho :\rho=(id\otimes \mathcal{N})\phi\}
		\end{equation}
	where the maximum is taken over all pure, bipartite states $\phi^{AA'}$.

	What makes it difficult to calculate the quantum capacity is the fact that, in general, the coherent information is non-additive, which leads to the fact that the quantum capacity can be non-additive (\cite{zero}):
		\begin{equation}
		\mathcal{Q}(\mathcal{N}_1\otimes\mathcal{N}_2)>\mathcal{Q}(\mathcal{N}_1)+\mathcal{Q}(\mathcal{N}_2)
		\end{equation}
	Therefore, we are having a look at a specific class of channels with some useful properties.

\section{Degradable channels and quantum capacity}

\begin{figure}[H]
					\centering
			  		\includegraphics[width=0.5\textwidth]{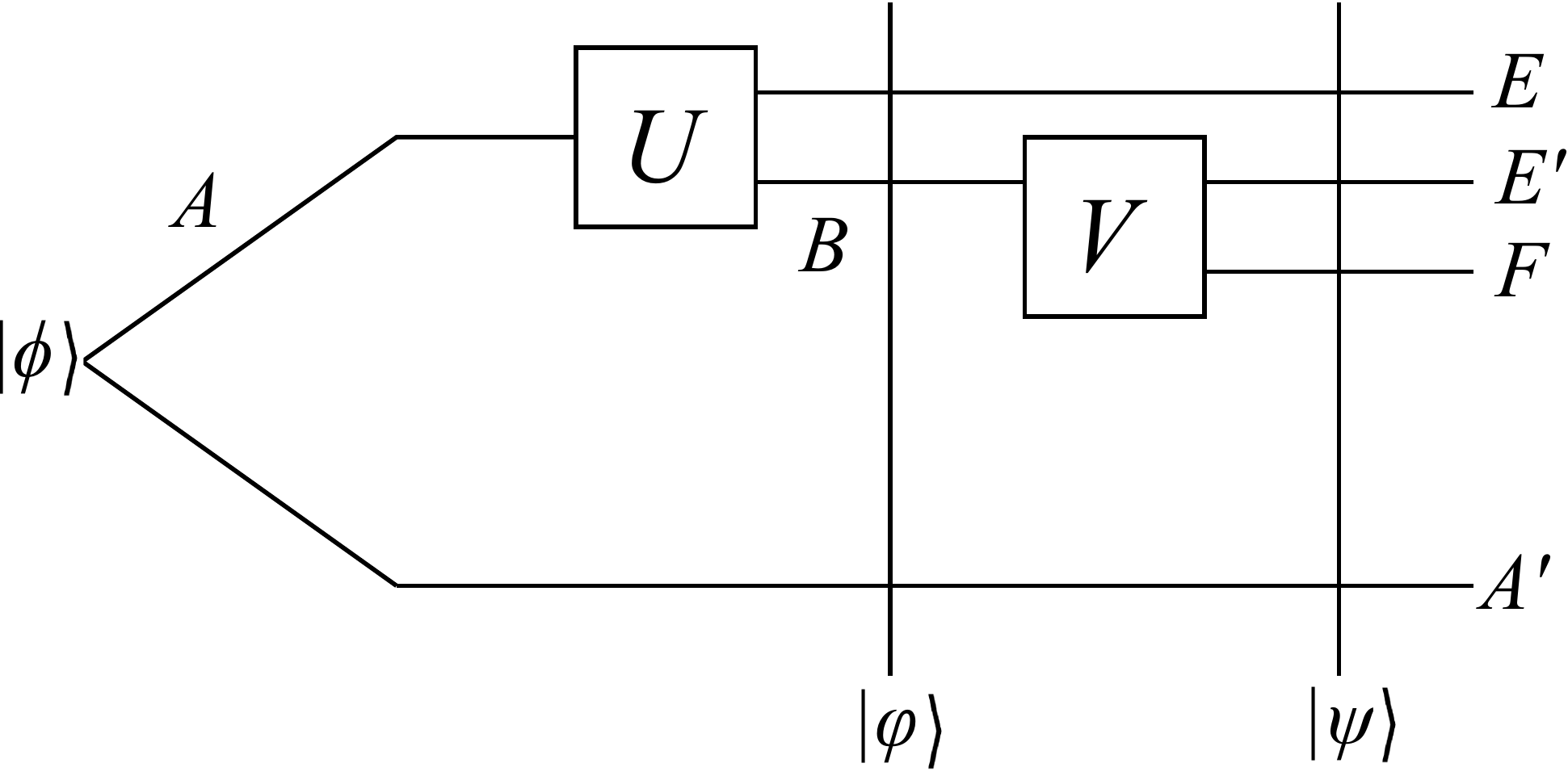}
					\caption{Schematic of a degradable quantum channel}
					\label{degr}
				\end{figure}

	We are especially interested in the quantum capacity of degradable channels since they are the only ones whose quantum capacity is known, which was shown in \cite{QCD}.
	\begin{D}[Degradable channel]
	A degradable quantum channel is one for which there exists a degrading map $\mathcal{T}^{B\rightarrow E}$ so that for any input state $\rho^{A}$:
		\begin{equation}
		(\mathcal{N}^c)^{A \rightarrow E }(\rho^{A})=\mathcal{T}^{B\rightarrow E}(\mathcal{N}^{A\rightarrow B}(\rho^{A}))
		\end{equation}
	\end{D}
	In brief, degradable channels are such that the receiver can simulate the channel to the environment by applying a degrading map to the channel output.
	Figure ~\ref{degr} shows a schematic of a degradable quantum channel. $\ket{\phi}$ is the input state, $\ket{\varphi}$ the output state and environment state and $\ket{\psi}$ the state shared between $A'$ (the reference system), $F$ and the two copies of the environment system, $E$ and $E'$ (see \cite{CMAW}).
	For degradable channels, the copies of the environment system, $E$ and $E'$ are the same, which leads to the fact that the coherent information becomes additive, which is a crucial requirement for the following theorem ~\ref{cohi}.

	\begin{theo}[Quantum capacity for degradable channels]
	For degradable channels, the quantum capacity is given by
		\begin{equation}
		\mathcal{Q}(\mathcal{N})=\max_{\rho^{AB}}I(A\rangle B)
		\end{equation}
	where $I(A\rangle B)$ is the coherent information
		\begin{equation}
		I(A\rangle B)_\rho=H(B)_\rho -H(AB)_\rho.
		\label{coh}
		\end{equation}
	\label{cohi}
	\end{theo}
	For the proof of this theorem, see \cite{QCD}. With this result, the quantum capacity is given by a single letter formula which makes it possible to calculate this quantity.

\section{Amplitude damping channel}

\begin{figure}[H]
				\centering
		  		\includegraphics[width=0.5\textwidth]{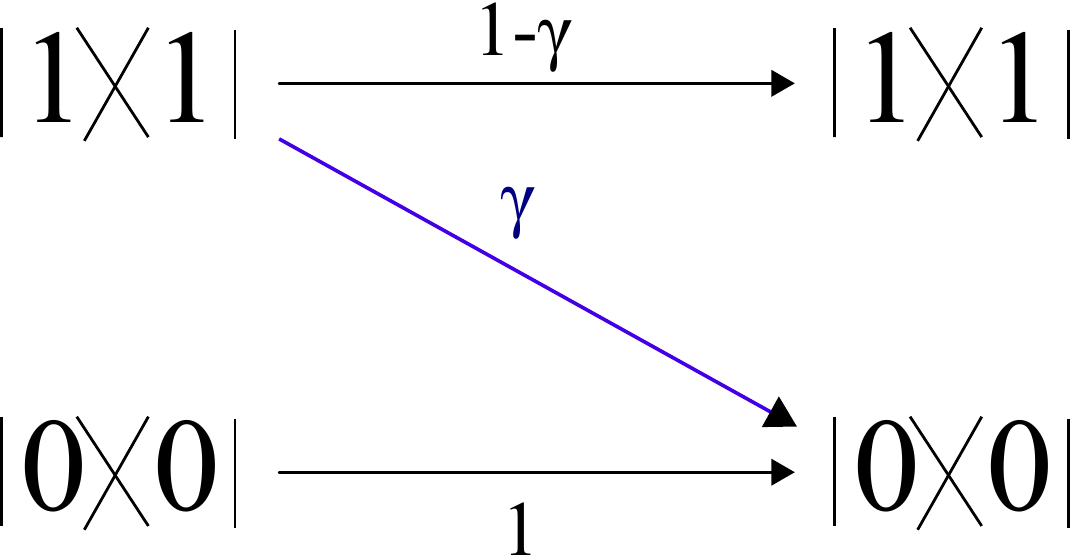}
				\caption{Interpretation of the qubit amplitude damping channel}
				\label{AD}
			\end{figure}

	One interesting channel is the amplitude damping channel.
	It models noisy quantum channels due to energy loss. One physical interpretation is thinking of the $\ket{0}$ state as the ground state of a two-level atom and the $\ket{1}$ state as the excited state of this atom. The amplitude damping channel models spontaneous emission, i.e. the transition from the excited state to the ground state that occurs with a probability of $0\le\gamma\le 1$, as shown in figure ~\ref{AD}.
	The amplitude damping channel can be described by the Kraus operators:
		\begin{equation}
		E_0=\begin{pmatrix}
		1 & 0 \\
		0 & \sqrt{1-\gamma}
		\end{pmatrix}
		\end{equation}
		\begin{equation}
		E_1=\begin{pmatrix}
		0 & \sqrt{\gamma} \\
		0 & 0
		\end{pmatrix}
		\end{equation}
	The channel acts on a state $\rho=\begin{pmatrix}
	a & b \\
	b^* & 1-a
	\end{pmatrix}$ in the following way:
		\begin{equation}
		\mathcal{E}_{AD}(\rho)=E_0\rho E_0^\dagger +E_1\rho E_1^\dagger=\begin{pmatrix}
		a+(1-a)\gamma & b\sqrt{1-\gamma} \\
		b^*\sqrt{1-\gamma} & (1-a)(1-\gamma)
		\end{pmatrix}
		\end{equation}

\section{Calculation}

\begin{figure}[htp]
					\centering
			  		\includegraphics[width=0.8\textwidth]{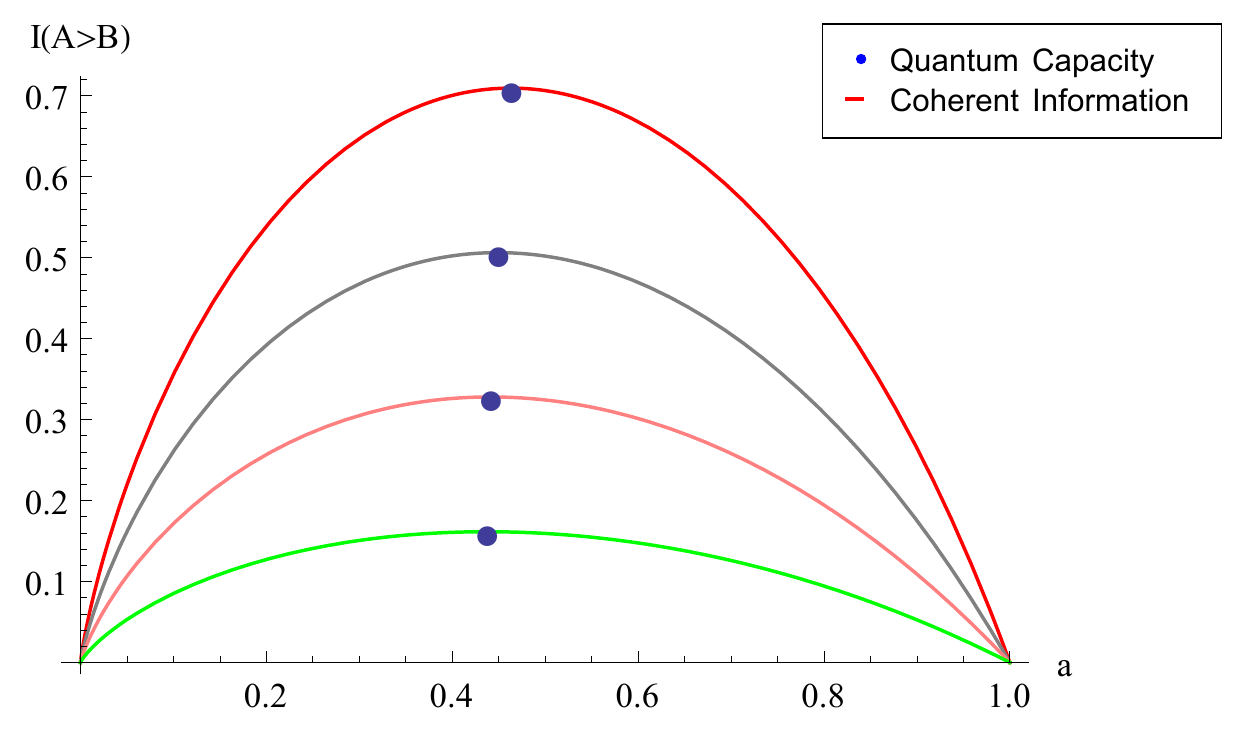}
					\caption{Coherent information with quantum capacity, for different $\gamma$}
					\label{QC2}
					\end{figure}

\begin{figure}[htp]
					\centering
			  		\includegraphics[width=0.8\textwidth]{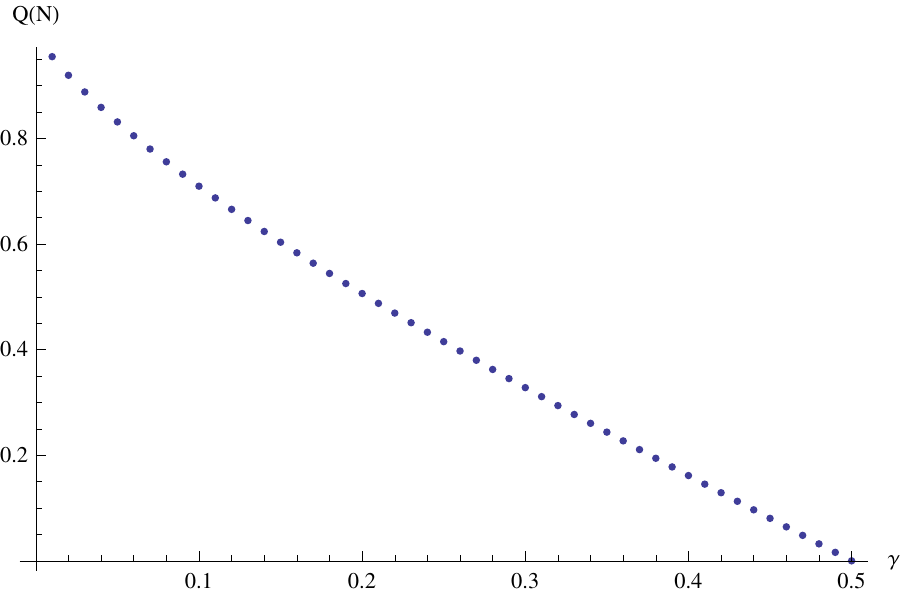}
					\caption{Quantum capacity $Q(N)$ as a function of $\gamma$ }
		        	\label{wild}
					\end{figure}

	To calculate the coherent information for the amplitude damping channel $\mathcal{E}_{AD}$, we use the following formula:
		\begin{align}
		\begin{split}
		I(A\rangle B)_\sigma &=H(B)_\sigma -H(AB)_\sigma\\
		&=H(\mathcal{E}_{AD}(\rho))-H(E)_\sigma\\
		&=H(\mathcal{E}_{AD}(\rho))-H(\mathcal{E}_{AD}^c(\rho))
		\end{split}
		\end{align}
	where $\rho$ is a pure input state
		\begin{equation}
		\rho=\begin{pmatrix}
		1-a & b^*\\
		b & a
		\end{pmatrix}
		\end{equation}
	and $\sigma$ is the state that Bob receives:
		\begin{equation}
		\sigma=\mathcal{E}_{AD}(\rho)=\begin{pmatrix}
				1-a(1-\gamma) & b\sqrt{1-\gamma} \\
				b^*\sqrt{1-\gamma} & a(1-\gamma)
				\end{pmatrix}.
		\end{equation}
	For the calculation, we also need to know how the complementary channel $\mathcal{E}_{AD}^c$ acts on $\rho$. Therefore, we use an isometric extension of the channel and then trace over Bob's system. The amplitude damping channel is of the form
		\begin{equation}
		\mathcal{E}_{AD}^{A\rightarrow B}(\rho^A)=\sum_j E_j \rho^A E_j^\dagger.
		\end{equation}
	and therefore, its isometric extension is given by the following map
		\begin{align}
		U_{\mathcal{E}_{AD}}^{A\rightarrow BE}&=\sum_j E_j\otimes \ket{j}^E\\
		&=\begin{pmatrix}
		0 & \sqrt{\gamma}\\
		1 & 0\\
		0 & 0\\
		0 & \sqrt{1-\gamma}.
		\end{pmatrix}
		\end{align}
	Now we apply this map to the input state $\rho$:
		\begin{equation}
		U_{\mathcal{E}_{AD}}^{A\rightarrow BE}(\rho)=\begin{pmatrix}
		a\gamma & b^*\sqrt{\gamma} & 0 & a\sqrt{1-\gamma}\sqrt{\gamma}\\
		b\sqrt{\gamma} & (1-a) & 0 & b\sqrt{1-\gamma}\\
		0 & 0 & 0 & 0 \\
		a\sqrt{1-\gamma}\sqrt{\gamma} & c\sqrt{1-\gamma} & 0 & a(1-\gamma)
		\end{pmatrix}.
		\end{equation}
	Tracing over Bob's system gives us the complementary channel:
		\begin{align}
		\mathcal{E}_{AD}^c(\rho)&=Tr_B(U_{\mathcal{E}_{AD}}^{A\rightarrow BE}(\rho))\\
		&=\begin{pmatrix}
		1-a\gamma & b^*\sqrt{\gamma} \\
		b\sqrt{\gamma} & a\gamma
		\end{pmatrix}
		\end{align}
	During the calculation, we will also need to calculate the entropy of the states,
		\begin{align}
		H(A)_\rho&=-Tr(\rho \log\rho)\\
		&=-\sum_i \lambda_i \log\lambda_i
		\end{align}
	where $\lambda_i$ are the eigenvalues of the state. These eigenvalues are
		\begin{gather}
		\lambda_{\pm_B}=\frac{1}{2}\left(1\pm \sqrt{((1+2a(\gamma-1))^2-4|b|^2(\gamma-1)}\right)\label{EVBob}\\
		\lambda_{\pm_E}=\frac{1}{2}\left(1\pm \sqrt{(1-2a\gamma)^2+4|b|^2\gamma}\right)\label{EVEve}.
		\end{gather}
	Currently, the coherent information is still a function of two variables, $a$ and $b$ (with fixed error parameter $\gamma$). It can be shown (as in \cite{MW}) that it is sufficient to consider only diagonal density operators to maximize the coherent information. Thus, the eigenvalues in ~\ref{EVBob} and ~\ref{EVEve} become
		\begin{gather}
		\lambda_{\pm_B}=\{(1-\gamma) a,1-(1-\gamma) a\}\label{Bobi}\\
		\lambda_{\pm_E}=\{\gamma a, 1-\gamma a\}\label{Evi}
		\end{gather}
	Now we have all the information we need to calculate the coherent information which is hence only a function of a. In figure ~\ref{wild}, this is diagrammed for different values of $\gamma$.
	\begin{figure}[htp]
					\centering
			  		\includegraphics[width=0.85\textwidth]{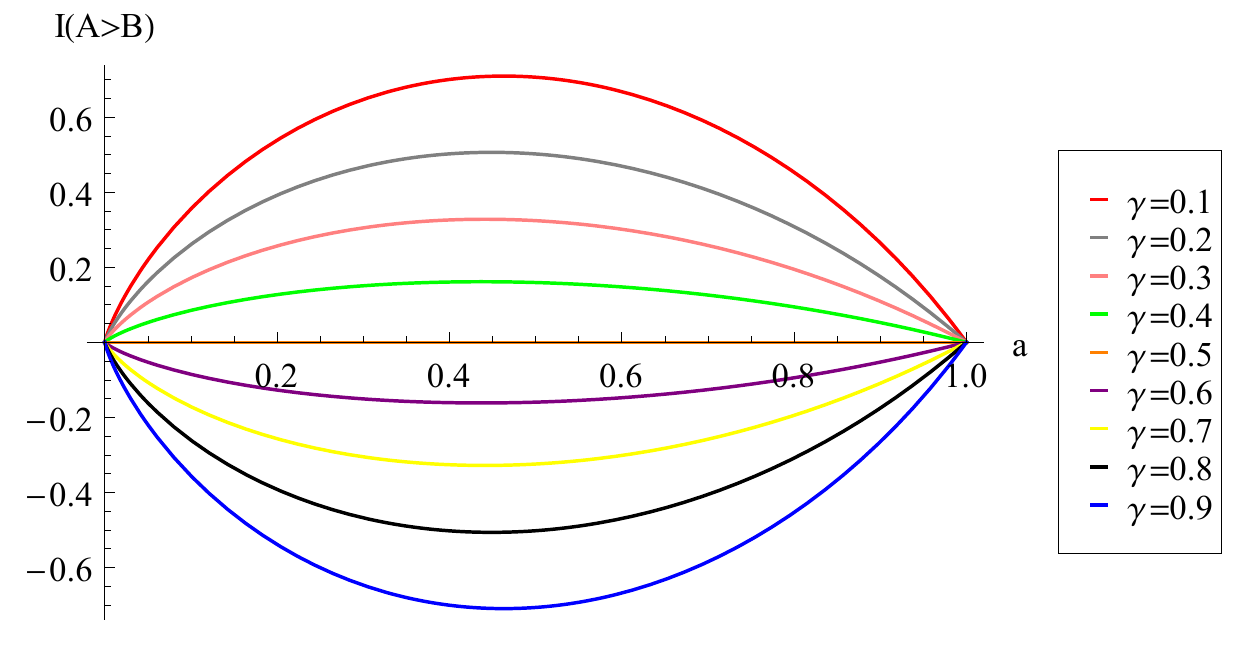}
					\caption{Coherent information as a function of $a$, for different $\gamma$}
					\label{QC1}
					\end{figure}
	For $\gamma<0.5$, the coherent information is concave and therefore a maximum exists. For $\gamma=0.5$ the coherent information vanishes, which can be easily understood since the eigenvalues of Bob (~\ref{Bobi}) and Eve (~\ref{Evi}) become the same. For $\gamma<0.5$, the coherent information is negative, which is clear because in that case Eve gets more information than Bob.\\
	\\
	In order to calculate the quantum capacity, we take the maximum of the coherent information over $a$. In figure ~\ref{QC2}, the coherent information with its particular maximum (the quantum capacity) is shown for several values of $0.1\le\gamma\le 0.5$.
	The coherent information can now be evaluated as a function of $\gamma$, which is shown in figure ~\ref{wild}. As expected, the quantum capacity decreases as the error parameter $\gamma$ increases.
	\newpage
	In \cite{AD}, the quantum capacity was calculated for Amplitude damping qubit channels and in \cite{small}, the quantum capacity has been calculated for all qubit channels that can be represented by two Kraus-Operators.\\
	\\
	The analysis of the quantum capacity leads to the question what happens if the key rate goes above this capacity, respectively if the strong converse theorem holds for the quantum capacity. Strong converse means that if the rate $r$ goes above the capacity $\mathcal{Q}(\mathcal{N})$, the error $\epsilon$ tends to $1$.\\
	\\
	It was proved in \cite{CMAW} for degradable channels that if the rate goes above the quantum capacity, the error jumps to $\frac{1}{\sqrt{2}}$ which is called the ''pretty`` strong converse. For a smaller class of channels (the generalized dephasing channels), the full strong converse was proved by
	\cite{strong}.

\chapter{Conclusion and outlook}

In this thesis, we presented the quantum key distribution protocol and gave an overview of the security proof as well as expressions for quantities like the final key length and the key rate. Furthermore, we examined a special class of channels, the degradable channels, and especially the amplitude damping channel, that models noise in a quantum channel. For this channel, the quantum capacity was calculated and diagrammed.\\
\\
Still, there are some open problems that are worth being investigated. We already mentioned the open tasks concerning quantum capacity. Since the strong converse theorem was only shown for degradable channels for an error of $0\le\epsilon\le\frac{1}{\sqrt{2}}$, it is open to show a full strong converse for degradable channels and later the strong converse for other types of channels.\\
\\
Another interesting task is to calculate the final key length that appears in theorem ~\ref{fivefive} or the (sifted) key rate in the non-asymptotic case as in lemma ~\ref{lem513}. These could be evaluated for different channels, for example the amplitude damping channel.



\bibliographystyle{alphadin}

\bibliography{Literatur}

\appendix
\chapter{Notation}
\begin{longtable}{ll}
$\perp$ & Abort symbol\\
$\not\perp$ & Passing symbol\\
$M$ & Number of states sent by Alice\\
$[M]$ & Set $\{1,2,...,M\}$\\
$n$ & Length of the raw key\\
$k$ & Length of the raw key used for parameter estimation\\
$l$ & Length of the final key\\
$r$ & key rate per signal\\
$r'$ & sifted key rate\\
$s$ & Length of the error correction syndrome\\
$t$ & Length of the hash used for verification in the error correcting scheme\\
$\delta$ & Threshold for the parameter estimation test\\
$\mathcal{N}_{A\rightarrow B}$ & Quantum channel between Alice and Bob\\
$\mathcal{M}_{A\rightarrow X|S}$ & Measurement map applied on register $A$ with setting $S$ and storing the result in $X$\\
$\mathcal{P}_{\emptyset\rightarrow A|RS}$ & Preparation map that returns a state in register $A$ depending on the setting $RS$\\
$M_{A_i}^{\phi,x}$ & Measurement operator acting on $A_i$ with the setting $\phi$ and outcome $x$\\
$c_i$ & Parameter quantifying the quality of the measurement on register $i$\\
$\bar{c}$ & Parameter quantifying the overall quality of the measurement\\
$A$ & Alice's initial quantum system\\
$B$ & Bob's initial quantum system\\
$E$ & Eve's system\\
$S^{\Phi_A}$ & Seed for the choice of Alice's measurement bases\\
$S^{\Phi_B}$ & Seed for the choice of Bob's measurement bases\\
$S$ & Register corresponding to all the seeds\\
$F^{pe}$ & Flag for the parameter estimation test\\
$F^{ec}$ & Flag for the error correction test\\
$F^{sift}$ & Flag for the sifting procedure\\
$F$ & Register corresponding to all the flags: $F=(F^{pe},F^{ec}, F^{sift})$\\
$C$ & Register containing all the transcripts\\
$R$ & Register of Alice's raw key\\
$T$ & Register for Bob's measurement results\\
$\Omega$ & Subset of $[M]$ for which Bob gets conclusive measurement results\\
$\Sigma$ & Subset of $m$ indices where Alice and Bob's settings agree

\end{longtable}

\clearpage


\end{document}